\documentclass[11pt]{article}
\usepackage{amstext}
\usepackage{amsfonts}
\usepackage{amsmath}

\newtheorem{theorem}{Theorem}[section]
\newtheorem{lemma}[theorem]{Lemma}

\def\square{\rule{2mm}{2mm}}

\newenvironment{proof}{{\noindent\bf Proof:  }}{\qquad\square}

\def\squarebox#1{\hbox to #1{\hfill\vbox to #1{\vfill}}}

\newcommand{\tensor}{\otimes}

\newcommand\meet\wedge

\newcommand{\adjoint}{\dagger}

\newcommand{\complex}{{\mathbb C}}

\newcommand{\trace}{{\rm Tr}}
\newcommand{\size}[1]{\left|#1\right|}

\newcommand{\floor}[1]{\left\lfloor#1\right\rfloor}
\newcommand{\ket}[1]{|#1\rangle}
\newcommand{\bra}[1]{\langle #1|}

\newcommand{\ketbra}[2]{\ket{#1}\!\bra{#2}}
\newcommand{\density}[1]{\ketbra{#1}{#1}}

\newcommand{\set}[1]{{\left\{#1\right\}}}

\newcommand{\identity}{{\mathbb I}}
\newcommand{\complexi}{{\mathrm i}}
\newcommand{\linear}{{\mathrm L}}

\newcommand{\comment}[1]{}
\newcommand{\suppress}[1]{}

\newcommand{\aitch}{{\mathcal{H}}}

\newcommand{\cI}{\mathcal{I}}
\newcommand{\cJ}{\mathcal{J}}
\newcommand{\cK}{\mathcal{K}}

\begin{document}

\title{\bf Invertible Quantum Operations and Perfect Encryption of
Quantum States}

\author{
Ashwin Nayak \\
U.\ Waterloo \& Perimeter \thanks{
Department of Combinatorics and Optimization, and Institute for
Quantum Computing, University of Waterloo, 200 University Ave.\ W.,
Waterloo, ON N2L 3G1, Canada.
E-mail: {\tt anayak@math.uwaterloo.ca}.
Research supported in part by NSERC, CIAR, MITACS, CFI, and OIT
(Canada).  A.N.\ is also Associate Member, Perimeter Institute for
Theoretical Physics, Waterloo, Canada. Research at Perimeter Institute
is supported in part by the Government of Canada through NSERC and by
the Province of Ontario through MEDT.
}
\and
Pranab Sen \\
TIFR \thanks{
School of Technology and Computer Science, Tata Institute of
Fundamental Research, Homi Bhabha Road, Colaba, Mumbai 400005, India.
Email: {\tt pgdsen@tcs.tifr.res.in}.
This research was done while the author was
at NEC Laboratories America, Inc.,
Princeton, NJ, U.S.A.  
}
}

\date{September~20, 2006}

\maketitle

\begin{abstract}
In this note, we characterize the form of an invertible quantum
operation, i.e., a completely positive trace preserving linear
transformation (a CPTP map) whose inverse is also a CPTP map. The
precise form of such maps becomes important in contexts such as
self-testing and encryption.  We show that these maps correspond to
applying a unitary transformation to the state along with an ancilla
initialized to a fixed state, which may be mixed.

The characterization of invertible quantum operations implies that
one-way schemes for encrypting quantum states using a classical
key may be slightly more
general than the ``private quantum channels'' studied by Ambainis,
Mosca, Tapp and de~Wolf~\cite[Section~3]{AmbainisMTW00}.  Nonetheless,
we show that their results, most notably a lower bound of~$2n$ bits of
key to encrypt~$n$ quantum bits, extend in a straightforward manner to
the general case.
\end{abstract}

\section{Introduction}
\label{sec-invertible}

The most general physically allowed operation on a quantum state
consists of coupling it to another system (an ancilla) in a known
state, via a unitary transformation, and then discarding part of the
system.  (In this article, a quantum state may be mixed and is
modeled by a density matrix.)  We say that a quantum operation~$E$ is
{\em invertible\/}, if there is another {\em quantum operation\/}~$D$
such that~$DE(\rho) = \rho$ for every state~$\rho$ in the domain
of~$E$.  Mathematically, a quantum operation corresponds to a {\em
completely positive trace preserving linear transformation\/}, a {\em
CPTP map\/}~\cite[Section~8.2]{NielsenC00}. A CPTP map may be an
invertible linear transformation, but may not correspond to an
invertible quantum operation in the sense defined above. For example,
the depolarizing channel is an invertible CPTP map, but its inverse is
not even positive. However, if a CPTP map corresponds to an invertible
quantum operation, it is necessarily also injective, and therefore 
invertible on its image.  This is because
there is a basis for the domain consisting of density matrices alone.
\comment{ Consider a basis~$\ketbra{i}{j}$ for~$\complex^d$. For
any~$i,j \in [d]$, $\density{i}$ are linearly independent. We can
generate~$\ketbra{i}{j}$ from the states~$\ket{i}, \ket{j}, \ket{i} +
\ket{j}, \ket{i} + \complexi \ket{j}$.  }

In this note, we characterize the form of an invertible quantum
operation. The precise form of such maps becomes important in contexts
such as self-testing~\cite{DamMMS00} and
encryption~\cite{AmbainisMTW00}.  A unitary operation is a natural
example of a CPTP map that is also invertible. It seems intuitively
obvious that {\em all\/} invertible quantum operations also be
unitary. This is indeed the case for CPTP maps transforming a Hilbert
space into itself~\cite[Chapter~3, Section~8,
Exercise~3.2]{Preskill98}. Here, we examine the more general case,
where an invertible quantum operation may take~$d$-dimensional states
to states in a Hilbert space of possibly larger dimension. We show
that these maps correspond to applying a unitary transformation to the
state along with an ancilla initialized to a fixed state, which may be
mixed (Theorem~\ref{thm-invertible}).  We also extend this
characterization to completely positive (CP) maps in
Theorem~\ref{thm-alt}. Its significance lies in the fact that when suitably
scaled by a positive real number, CP maps correspond to the result
of getting one of a subset of outcomes on a measurement.

Invertible quantum operations also occur in the context of error
correction. There, the goal is to find a linear subspace of a Hilbert
space such that the restriction of the noise operator to this space is
invertible. The proof of our characterization theorem closely follows
the proof of the error-correction criterion~\cite[page~436,
Theorem~10.1]{NielsenC00}. 

A notion related to error-correction is that of a {\em reversible
quantum operation\/}. Several authors~\cite{NielsenC97,NielsenCSB98}
consider operations that are completely positive maps defined by the
process of making a measurement, and getting one of a subset of
outcomes. They call such an operation~$E$ {\em reversible\/} on a
subspace, if for all states~$\rho$ in the subspace, there is a quantum
operation~$D$ such that~$D(E(\rho)) / \trace(E(\rho)) = \rho$. Nielsen
{\em et al.\/}~\cite{NielsenCSB98} characterize such operations in
information theoretic as well as algebraic terms (akin to the
error-correction criterion).

Theorem~\ref{thm-invertible} has implications for perfect encryption
of quantum states using a classical private key (see, e.g.,
Ref.~\cite{AmbainisMTW00}). These protocols for encryption, also
called ``private quantum channels'' by some authors, involve two
parties, labeled Alice and Bob. The two parties share a secret,
uniformly random bit-string~$k$, called the private key. Alice wishes
to send a quantum message, a~$d$-dimensional quantum state~$\rho$, to
Bob. She would like to apply an invertible CPTP map~$E_k$ to the
state, and send it to Bob so that when averaged over~$k$, the result
is a fixed density matrix (independent of the message). This would
ensure that no eavesdropper be able to distinguish two different
messages with any degree of success, and therefore guarantee
information theoretic security.  Bob, who also has the key~$k$, can
apply the inverse operation~$D_k$ to decrypt the message~$\rho$
perfectly. (We have implicitly assumed that the quantum channel is
noiseless unless an eavesdropper tampers with it.)

The characterization of invertible quantum operations implies that in
the most general one-way encryption scheme, Alice may apply a unitary
operation to the state to be encrypted along with an ancilla {\em that
depends upon the key\/}. This is slightly more general than the form
studied by Ambainis, Mosca, Tapp and
de~Wolf~\cite[Section~3]{AmbainisMTW00}, where the ancilla is assumed
to be independent of the key.  Nonetheless, their results, most
notably a lower bound of~$2n$ bits of key to encrypt~$n$ quantum bits,
extend in a straightforward manner to the general case. We summarize
these observations in Section~\ref{sec-encryption}. 

The lower bound of~$2n$ classical key bits needed to encrypt quantum
states was also shown by Boykin and Roychowdhury~\cite{BoykinR03},
{\em assuming that no ancilla is used\/}.  Their proof was simplified
by Ambainis {\it et al.\/}~\cite{AmbainisMTW00}. Since the addition of
ancilla results in longer ciphertext, and hence is less efficient,
this case is of special interest.  We observe that this~$2n$ lower
bound follows directly from a ``rank argument''.

We point out that information theoretic proofs due to
DiVincenzo, Hayden, and Terhal~\cite[Section IV]{DiVincenzoHT03} and 
Jain~\cite{Jain05} follow a different route to the same lower bound
on the length of key for general one-way encryption schemes. We also
note that the requirement of perfect information theoretic security
imposes an additional constraint on the maps~$E_k$, apart from
invertibility. This constraint may simplify the mathematical structure
of these schemes, and further simplify the proofs we give.

\section{Invertible quantum operations}
\label{sec-thm}

We refer the reader to the text~\cite{NielsenC00} for basic concepts
related to quantum states and operations, and present our
characterization theorem directly.

Let~$\linear(\aitch)$ denote the set of linear operators on the
Hilbert space~$\aitch$. 

\begin{theorem}
\label{thm-invertible}
Let~$E : \linear(\complex^p) \rightarrow \linear(\complex^q)$ be a
completely positive, trace preserving linear transformation (a CPTP map). Suppose
there is a CPTP map~$D : \linear(\complex^q) \rightarrow
\linear(\complex^p)$ such that~$DE(\rho) = \rho$ for all density
matrices~$\rho \in \linear(\complex^p)$. I.e., $E$ is an invertible
quantum operation with inverse~$D$.

Then there is a density matrix~$\omega \in
\linear(\complex^{\floor{q/p}})$, and a unitary operation
on~$\complex^q$ such that~$E(\rho) = U (\rho \tensor \omega)
U^\adjoint$. Furthermore, $D$ corresponds to applying~$U^\adjoint$,
and tracing out the~$\floor{q/p}$ dimensional ancilla.
\end{theorem}

\begin{proof} 
%(of Theorem~\ref{thm-inv})
As mentioned in Section~\ref{sec-invertible}, there is a close analogy
between error-correction, and the invertibility of quantum operations.
If we view~$\complex^p$ as a code subspace, and~$E$ as a noisy channel
restricted to this subspace, then the decoding map~$D$ corrects
any ``errors'' introduced by~$E$. We may thus appeal to the
error-correction criterion~\cite[page~436, Theorem~10.1]{NielsenC00}
to give a short proof of the theorem. Instead, in the interest of
completeness, we present the details below. Those familiar with the
criterion may skip to Equation~(\ref{eqn-cond}) and then to
Equation~(\ref{eqn-orth}) after picking up the notation in the next
paragraph.

A CPTP map from~$\linear(\complex^n)$ to~$\linear(\complex^m)$ can be
expressed in terms of linear transformations from~$\complex^n$ to~$\complex^m$
({\em Kraus operators\/}), using an operator sum
representation~\cite[Exercise~8.3]{NielsenC00}. Suppose we express
both maps~$E$ and~$D$ in terms of some set of Kraus
operators~$\set{A_i}_{i \in \cI}$ and~$\set{B_j}_{j \in \cJ}$, respectively.
Then, by the
invertibility of~$E$, we have, for every~$\rho \in \linear(\complex^p)$
\begin{eqnarray*}
DE(\rho) \quad = \quad 
\sum_{i \in \cI} \sum_{j \in \cJ} 
B_j A_i \; \rho \; A_i^\adjoint B_j^\adjoint & = & \rho.
\end{eqnarray*}
Thus, the CPTP operation~$DE$ defined on~$\linear(\complex^p)$ may
equivalently be expressed in terms of the single Kraus 
operator~$\identity_p$,
the identity operator on~$\complex^p$. By the unitary equivalence of
Kraus representations~\cite[Page~372, Theorem~8.2]{NielsenC00}, there
are complex numbers~$\alpha_{ij}$, such that~$\sum_{i \in \cI}
\sum_{j \in \cJ} \size{\alpha_{ji}}^2 = 1$, and for all $i \in \cI$,
$j \in \cJ$,
\begin{eqnarray*}
B_j A_i & = & \alpha_{ji}\; \identity_p \, .
\end{eqnarray*}
Therefore for all $i, i' \in \cI$,
\begin{eqnarray*}
\sum_{j \in \cJ} A_{i'}^\adjoint B_j^\adjoint B_j A_i 
    & = & \beta_{i' i} \; \identity_p,
\end{eqnarray*}
where
\begin{eqnarray}
\label{eqn-beta}
\beta_{i'i} \quad = \quad \sum_{j \in \cJ} \bar{\alpha}_{ji'} \alpha_{ji}.
\end{eqnarray}
Observe that $M = (\beta_{i'i})_{i',i \in \cI}$ is a Hermitian matrix.
Since, $\sum_{j \in \cJ} B_j^\adjoint B_j = \identity_q$, we have
for all $i, i' \in \cI$,
\begin{eqnarray}
\label{eqn-cond}
A_{i'}^\adjoint A_i & = & \beta_{i'i} \identity_p.
\end{eqnarray}
The conditions in Equation~(\ref{eqn-cond}) imply that each Kraus
operator~$A_i$ is a scaled isometric embedding of~$\complex^p$
into~$\complex^q$. However, the resulting images need not be mutually
orthogonal. We therefore first derive an equivalent representation
for~$E$ in which the Kraus operators embed into orthogonal subspaces.

Using equation~(\ref{eqn-beta}) above, we have that for any 
vector~$x \in \complex^{\cI}$,
\begin{eqnarray*}
(x^\adjoint M x) 
    & = & \sum_{i,i' \in \cI} \bar{x}_{i'} \, \beta_{i'i} \, x_i \\
    & = & \sum_{i,i' \in \cI} \sum_{j \in \cJ} \bar{x}_{i'} \,
          \bar{\alpha}_{ji'} \; \alpha_{ji} \, x_i \\
    & = & \sum_{j \in \cJ} 
          \left( \sum_{i' \in \cI} \bar{x}_{i'} \, \bar{\alpha}_{j i'} \right)
          \left( \sum_{i \in \cI} x_i \, \alpha_{j i} \right) \\
    & \geq & 0.
\end{eqnarray*}
So the matrix~$M$ is positive semi-definite.  Moreover
%\begin{eqnarray*}
\[
\trace(M) \quad = \quad \sum_{i \in \cI} \beta_{ii} 
    \quad = \quad \sum_{i \in \cI} \sum_{j \in \cJ} 
                  \size{\alpha_{ji}}^2 \quad = \quad 1.
\]
%\end{eqnarray*}

Let~$V = (v_{i'i})_{i',i \in \cI}$ be a unitary matrix that
diagonalizes~$M$. Let~$\Gamma = V^\adjoint M V$ be the resulting
diagonal matrix, with~$\gamma_i = \Gamma_{ii} \geq 0$ for all $i \in
\cI$, and~$\sum_{i \in \cI} \gamma_i = \sum_{i \in \cI} \beta_{ii} =
1$.  Then the Kraus operators~$C_k = \sum_i v_{ik} \, A_i$, $k \in \cI$
also represent the same map~$E$, as may be checked by direct
substitution (cf.~\cite[Page~372, Theorem~8.2]{NielsenC00}).
Moreover, the range spaces of the various operators~$C_k$ are
orthogonal. In fact for $k, k' \in \cI$,
\begin{eqnarray}
C_{k'}^\adjoint C_k 
    & = & \left( \sum_{i' \in \cI} \bar{v}_{i'k'} \, A_{i'}^\adjoint \right) 
          \left( \sum_{i \in \cI} v_{ik}\, A_i \right) 
          \nonumber \\
    & = & \sum_{i', i \in \cI}  \bar{v}_{i'k'} \, v_{ik} 
          \left( A_{i'}^\adjoint A_i \right) \nonumber \\
    & = & \sum_{i', i \in \cI}  \bar{v}_{i'k'} \, v_{ik} 
          \left( \beta_{i'i} \, \identity_p \right) \nonumber \\
    & = & \left( \sum_{i', i \in \cI}  \bar{v}_{i'k'} \, \beta_{i'i} \, v_{ik} 
          \right) \cdot \identity_p \nonumber \\ 
    & = & \Gamma_{k' k} \, \identity_p \nonumber \\
    & = & \delta_{k' k} \, \gamma_k \, \identity_p,  \label{eqn-orth}
\end{eqnarray}
where~$\delta$ is the Kronecker delta function. 

Define $\cK = \set{k \in \cI: \gamma_k \neq 0}$.
Looking at the singular value decomposition of~$C_k$, $k \in \cK$, we 
now conclude
that all its singular values are equal to~$\sqrt{\gamma_k}$, and
that the various operators~$C_k$ are scaled unitary embeddings
of~$\complex^p$ into 
orthogonal subspaces of~$\complex^q$: $C_k =
\sqrt{\gamma_k} \; \sum_{l \in [p]} \; \ketbra{y_{kl}}{u_l^{(k)}}$,
where~$\set{y_{kl}}_{k \in \cK, l \in [p]}$ is an orthonormal set of
vectors in $\complex^q$, and~$\set{u_l^{(k)}}_{l \in [p]}$
is an orthonormal basis of $\complex^p$ for each~$k \in \cK$. 
As a consequence, $q \geq p$, and~$\size{\cK} \leq \floor{q/p}$.

We may now define~$\omega = \sum_{k \in \cK} \; \gamma_k \; \density{w_k}$,
where~$\set{w_k}_{k \in \cK}$ is an orthonormal set
in~$\complex^{\floor{q/p}}$.  Define~$U$ as any unitary extension to
$\complex^q$ of the map:
\begin{eqnarray*}
\ket{u_l^{(k)}} \tensor \ket{w_k} & \mapsto & \ket{y_{kl}},
\end{eqnarray*}
where $k \in \cK$ and $l \in [p]$.
A straightforward check confirms that~$E$ may be implemented by
applying ~$U$ to any state in~$\complex^p$ tensored with
ancilla~$\omega$. Similarly, the inverse operation~$D$ may be
implemented by applying~$U^\adjoint$ and tracing out the
state~$\omega$.
\end{proof}

The proof of the invertibility criterion tells us how to deal with the
subtlety that the ancillary density matrix~$\omega$ may be expressed
as a multitude of mixtures, each of which gives rise to a different
Kraus representation for the map~$E$. For an arbitrary
mixture~$\sum_t \; r_t \, \density{\phi_t} = \omega$, the resulting Kraus
operators~$E_t = \sqrt{r_t} \, U (\identity_p \tensor \ket{\phi_t})$ are
not necessarily in the form from which the operator~$U$ is
evident. The diagonalization of the matrix~$M$ in the proof
corresponds exactly to the diagonalization of~$\omega$ and this allows
us to ``read out'' the unitary matrix, and the ancilla state~$\omega$
itself.

The converse of our theorem is manifestly true, so it provides a
characterization of invertible quantum operations.
An alternative characterization was
pointed out to us by Jon Tyson~\cite{Tyson06}. 
Below, we state an extension
of his characterization to completely positive (CP) but not necessarily
trace preserving maps, and sketch its proof.
\begin{theorem}
\label{thm-alt}
A completely positive (CP) linear 
transformation~$E : \linear(\complex^p)
\rightarrow \linear(\complex^q)$ has a CP inverse $D$ iff there exists
a positive semi-definite linear operator $Q \in \linear(\complex^q)$ and
a real number~$c > 0$ such that
for all density matrices~$\rho,\sigma \in \linear(\complex^p)$,
\[
\trace(Q \, E(\rho) \, Q \, E(\sigma)) \quad = \quad c \cdot
\trace(\rho \sigma).
\]
In addition, a CPTP map $E$ has a CPTP inverse $D$ iff 
$Q$ may be taken to be the identity operator $\identity_q$ in the above
characterization.
\end{theorem}
\begin{proof}{
We first sketch a proof of the forward direction of the theorem.
Suppose we have a CP map $E$ with a CP inverse $D$.
Express~$E$ and~$D$ in terms of some set of Kraus
operators~$\set{A_i}_{i \in \cI}$ and~$\set{B_j}_{j \in \cJ}$, respectively.
Arguing as in the proof of Equation~(\ref{eqn-cond}) of
Theorem~\ref{thm-invertible}, we have for all $i, i' \in \cI$,
\begin{eqnarray}
\label{eqn-newcond}
A_{i'}^\adjoint Q A_i & = & \beta_{i'i} \; \identity_p,
\end{eqnarray}
where $Q = \sum_{j \in \cJ} B_j^\adjoint B_j$ and $\beta_{i'i}$ is as in
Equation~(\ref{eqn-beta}). Note that $Q$ is positive
semi-definite. From Equation~(\ref{eqn-newcond}), it 
follows that~$\trace(Q \, E(\rho) \, Q \, E(\sigma)) 
= c \cdot \trace(\rho \sigma)$, where~$c = \sum_{i,i' \in \cI} 
|\beta_{i,i'}|^2$. Note that~$c > 0$ since~$\sum_{i \in \cI} 
\beta_{ii} = 1$ as in the proof of Theorem~\ref{thm-invertible}.

We now sketch a proof of  the reverse direction of the theorem.
The condition 
$\trace(Q \, E(\rho) \, Q \, E(\sigma)) = c \cdot \trace(\rho \sigma)$,
with~$Q$ positive semi-definite and $c > 0$ implies that
\begin{eqnarray*}
c \; \trace(\rho \sigma) 
    & = & \trace(Q \, E(\rho) \, Q \, E(\sigma)) 
\suppress{
      =  
\trace
\left(\sum_{i \in \cI} Q A_i \rho A_i^\adjoint\right)
\left(\sum_{i' \in \cI} Q A_{i'} \sigma A_{i'}^\adjoint \right) \\
}
    \quad = \quad  \trace \left( \sum_{i,i' \in \cI}
                   Q A_i \rho A_i^\adjoint
                   Q A_{i'} \sigma A_{i'}^\adjoint \right) \\
    & =  & \trace \left( \rho \sum_{i,i' \in \cI}
           A_i^\adjoint Q A_{i'} \sigma A_{i'}^\adjoint Q A_i
           \right),
\end{eqnarray*}
for all density matrices $\rho$, $\sigma$. This means that
$
\sum_{i,i' \in \cI} A_i^\adjoint Q  A_{i'} \sigma A_{i'}^\adjoint Q A_i = 
c \, \sigma
$
for all density matrices $\sigma$, since density
matrices $\rho$ form a basis for $\linear(\complex^p)$. This implies
that the CPTP map defined by 
$\set{c^{-1/2} A_i^\adjoint Q A_{i'}}_{i,i' \in \cI}$
is the identity map on $\linear(\complex^p)$. 
By the unitary equivalence of Kraus operators,
we see that there are complex numbers
$\set{\beta_{i,i'}}_{i,i' \in \cI}$, $\sum_{i,i'} |\beta_{i,i'}|^2 = c$ 
such that 
$A_i^\adjoint Q A_{i'} = \beta_{i,i'} \identity_p$, 
for all $i,i' \in \cI$. 
Taking trace, we have
$\beta_{i,i'} = p^{-1} \trace A_i^\adjoint Q A_{i'}$.
The matrix $M = (\beta_{i,i'})_{i,i' \in \cI}$
is Hermitian, and by arguing as in the proof of
Theorem~\ref{thm-invertible} we see that $M$ is positive semi-definite.
Let $E'$ denote the CP map given by the
Kraus operators $\set{Q^{1/2} A_i}_{i \in \cI}$. 
Now we follow the proof of Theorem~\ref{thm-invertible} 
from the argument for
Equation~(\ref{eqn-orth}) onwards to conclude that there exists 
a positive semi-definite matrix~$\omega \in
\linear(\complex^{\floor{q/p}})$, and a unitary operation $U$
on~$\complex^q$ such that~$E'(\rho) = U (\rho \tensor \omega )
U^\adjoint$ for all density matrices $\rho$ in $\linear(\complex^p)$. 
This shows that $E$ has a CP inverse $D$ 
which corresponds to conjugating by~$U^\adjoint Q^{1/2}$,
and tracing out the~$\floor{q/p}$-dimensional ancilla $\omega$.

Suppose now that CPTP map $E$ has a CPTP inverse $D$. 
In the above argument, we get 
$Q = \sum_{j \in \cJ} B^\adjoint_j B_j = \identity_q$.
Conversely, suppose that CPTP map $E$ satisfies
$\trace(E(\rho) E(\sigma)) = c \cdot \trace(\rho \sigma)$, $c > 0$
for all density matrices $\rho, \sigma \in \linear(\complex^p)$.
The trace preserving property of $E$ implies that the matrix $M$ in the
above argument has unit trace. This implies that the positive semi-definite
matrix $\omega$ has unit trace, that is, $\omega$ is a density matrix.
This shows that $E$ has a CPTP inverse $D$ 
which corresponds to applying~$U^\adjoint$,
and tracing out the~$\floor{q/p}$-dimensional ancilla $\omega$.
}
\end{proof}

\section{Perfect encryption of quantum states}
\label{sec-encryption}

A one-way protocol for perfect encryption of quantum states
in~$\linear(\complex^d)$ consists of a probability
distribution~$\set{p_k, E_k}$ over invertible quantum operations~$E_k
: \linear(\complex^d) \rightarrow \linear(\complex^D)$ such that the
image of every state~$\rho \in \linear(\complex^d)$ under the map
\begin{eqnarray*}
R(\rho) & = & \sum_k p_k \, E_k(\rho)
\end{eqnarray*}
is a fixed state~$\sigma \in \linear(\complex^D)$.  This is also known
as a {\em randomization scheme\/}, or a {\em private quantum
channel\/}.

As mentioned in Section~\ref{sec-invertible}, the probability
distribution~$\set{p_k}$ corresponds to a random secret key that two
parties Alice and Bob share. The map~$E_k$ is an encryption map that
Alice applies to her quantum message~$\rho$, and its inverse is the
decryption map that Bob applies to retrieve the message. To an
eavesdropper with no information about the secret key, the density
matrix of the ciphertext is exactly~$\sigma = R(\rho)$. Since this is
completely independent of the message, the protocol achieves
information theoretic security.

Our characterization theorem from the previous section implies that
the most general one-way quantum encryption scheme~$R$ (with no
decoding error in the absence of eavesdropping) is of the following
form: for each value of key~$k$, there is an ancilla~$\omega_k$,
possibly mixed, and a unitary~$U_k$ such that~$E(\rho) = \sum_k p_k \,
U_k (\rho \tensor \omega_k) U_k^\adjoint$. This is slightly more
general than the form assumed by Ambainis, Mosca, Tapp, and
de~Wolf~\cite[Section~3]{AmbainisMTW00}, in that the ancilla may
depend on the value of the key.  However, their results, especially a
proof that~$2n$ bits of key are required to encrypt~$n$ quantum bits,
extend to this form of encryption in a straightforward manner. Below
we give a sketch of this extension.

We begin with the following the lemma.
\begin{lemma}
\label{thm-orthogonal}
Let~$\set{p_k, E_k}$ define a perfect encryption map~$R$
for~$d$-dimensional quantum states in~$\linear(\complex^d)$. Then, for
any two orthogonal states~$\ket{i}, \ket{j} \in \linear(\complex^d)$,
$R(\ketbra{i}{j}) = 0$.
\end{lemma}
A simple proof of this lemma occurs in Theorem~5.2 of
Ref.~\cite{MoscaTW00}, and works verbatim for an encryption scheme as
described above. We need only consider the action of~$R$ on the
states~$\ket{i}, \ket{j}, \frac{1}{\sqrt{2}} (\ket{i} + \ket{j}),
\frac{1}{\sqrt{2}} (\ket{i} + \complexi \ket{j})$, where~$\complexi =
\sqrt{-1}$, to arrive at the lemma.  (A stronger version of this lemma
occurs as Lemma~4.4 in Ref.~\cite{AmbainisMTW00} and also generalizes
verbatim.)

An immediate corollary of Lemma~\ref{thm-orthogonal} is that if one
half of a bipartite Bell state is encrypted, the resulting bipartite
state is independent of which Bell state was encrypted. In fact, if
the encryption procedure is applied to the first half of the {\em
any\/} input Bell state, the resulting state is proportional
to~$\sigma \otimes \identity$ where $\sigma$ the output state of the
encryption procedure.  Using this property, Ambainis {\em et
al.\/}~\cite{MoscaTW00,AmbainisMTW00} show that any protocol to
encrypt~$n$ quantum bits may be transformed to a protocol that
encrypts~$2n$ classical bits.
\begin{lemma}
\label{thm-classical}
Let~$\set{p_k, E_k}$ define a perfect encryption map~$R$ for~$n$ qubit
states. Then, there is a map~$R'$ given by a distribution~$\set{p_k,
E_k'}$ that perfectly encrypts~$2n$ classical bits (i.e., a fixed
basis of~$\complex^{2^{2n}}$).
\end{lemma}
The idea behind this lemma is to encode the~$2n$ bits into orthogonal
Bell states over~$2n$ qubits, then encrypt one half of the Bell state
using~$R$ and finally send the bipartite state across as the encrypted
message. The map~$R'$ is given by the composition of these steps. 

Finally, we show how to extend Theorem~5.3 of
Refs.~\cite{MoscaTW00,AmbainisMTW00}. The proof relies on concepts
from quantum information theory. We refer the reader to the two
papers, and the text~\cite{NielsenC00} for the required background.
\begin{lemma}
\label{thm-entropy}
Let~$\set{p_k, E_k}$ define a perfect encryption map~$R$ for~$m$
classical bits. Then, the Shannon entropy~$H(p)$ of the
distribution~$p$ is at least~$m$.
\end{lemma}
\begin{proof}
Consider~$\sigma$, the result of encrypting the basis state~$
\density{0}$. Then,~$R(\density{0}) = \sigma = R(\identity/2^m)$,
since the completely mixed state may be viewed as a mixture of
(classical) basis states. So
\begin{eqnarray*}
\sigma & = & \sum_k p_k \, E_k(\density{0}) \\
    & = & \sum_k p_k \, U_k (\density{0} \tensor \omega_k) U_k^\adjoint \\
    & = & \sum_k p_k  \, U_k \left( \frac{\identity}{2^m} 
          \tensor \omega_k \right) U_k^\adjoint .
\end{eqnarray*}
Invoking Theorem~11.10 on page~518 of Ref.~\cite{NielsenC00}, the von
Neumann entropy of~$\sigma$ may be bounded above as
\begin{eqnarray*}
S(\sigma) & = & S \! \left( \sum_k p_k \, E_k(\density{0}) \right) \\
    & \leq & H(p) + \sum_k p_k \; S(E_k(\density{0})) \\
    & = & H(p) + \sum_k p_k \; 
          S(U_k (\density{0} \tensor \omega_k) U_k^\adjoint) \\
    & = &  H(p) + \sum_k p_k \; S(\omega_k).
\end{eqnarray*}
By concavity of von Neumann entropy, $S(\sigma)$ may also be bounded
from below as
\begin{eqnarray*}
S(\sigma) & = & S \! \left( \sum_k p_k  \, U_k ( \frac{\identity}{2^m} 
          \tensor \omega_k) U_k^\adjoint \right) \\
    & \geq & \sum_k p_k \; S \! \left(  U_k ( \frac{\identity}{2^m} 
          \tensor \omega_k) U_k^\adjoint \right) \\
    & = & \sum_k p_k \; 
          S \! \left( \frac{\identity}{2^m} \tensor \omega_k  \right) \\
    & = & \sum_k p_k \; \left( S(\identity/2^m) + S(\omega_k) \right) \\
    & = & m + \sum_k  p_k \;  S(\omega_k).
\end{eqnarray*}
The two bounds together give~$H(p) \geq m$.
\end{proof}

Lemmas~\ref{thm-classical} and~\ref{thm-entropy} imply:
\begin{theorem}
\label{thm-lb}
Let~$\set{p_k, E_k}$ define a perfect encryption map~$R$ for~$n$
qubits.  Then, the Shannon entropy~$H(p)$ of the distribution~$p$ is
at least~$2n$.
\end{theorem}

A weaker version of this theorem, where the encryption
operations~$E_k$ are chosen to be unitary, was shown by Boykin and
Roychowdhury~\cite{BoykinR03}. We sketch how in this case, a lower
bound of~$2n$ bits for the size of key follows from a simple rank
argument. 

Let~$\set{p_k, U_k}$ define a perfect encryption map~$R$ for~$n$
qubits for some unitary operators~$U_k$.  Note that~$R$ is a unital
map; it maps the completely mixed state to itself. Therefore, the
output state of~$R$ is the completely mixed state~$\sigma =
\frac{\identity_{2^n}}{2^n}$. For any bipartite pure state~$\rho$
on~$2n$ qubits, the rank of~$(\identity \tensor R)\rho$, where~$R$
acts on one half of~$\rho$, is at most the number of non-zero~$p_k$.
However, from the corollary to Lemma~\ref{thm-orthogonal} mentioned
above, if we choose~$\rho$ to be any pure bipartite Bell state, one
half of which is encrypted,
\[
(\identity \tensor R)\rho 
    \quad = \quad \frac{\identity_{2^n}}{2^n} \otimes \sigma 
    \quad = \quad \frac{\identity_{2^{2n}}}{2^{2n}},
\]
which has rank~$2^{2n}$.  Thus, the support of the probability
distribution of the secret key has size at least~$2^{2n}$, which gives
the claimed lower bound. Note that this does not imply the stronger
claim of Theorem~\ref{thm-lb} that the entropy of the distribution
is~$2n$, or the stronger characterization of optimal perfect
encryption schemes (without ancilla) due to Boykin and
Roychowdhury~\cite[Section III]{BoykinR03}.

%\section*{Acknowledgements}

%\bibliographystyle{plain}
%\bibliography{crypto,books}

\end{document}